\documentclass[sigconf,screen]{acmart}
\AtBeginDocument{%
  }

\copyrightyear{2026}
\acmYear{2026}
\setcopyright{cc}
\setcctype{by}
\acmConference[ASE '26]{Proceedings of the 41st IEEE/ACM International Conference on Automated Software Engineering}{October 12--16, 2026}{Munich, Germany}
\acmBooktitle{Proceedings of the 41st IEEE/ACM International Conference on Automated Software Engineering (ASE '26), October 12--16, 2026, Munich, Germany}
\acmDOI{10.1145/3832783.3834355}
\acmISBN{979-8-4007-2882-2/2026/10}

\newcommand{\name}{\textsc{AgentExecutor}}

\newcommand{\phead}[1]{\vspace{1mm} \noindent {\bf #1}}

\usepackage[ruled,vlined]{algorithm2e}
\usepackage{booktabs}
\usepackage{longtable}
\usepackage{graphicx}
\usepackage{tcolorbox}
\usepackage{minted}
\usepackage{pifont}
\usepackage{enumitem}
\usepackage{pbalance}
\newcommand{\greyboxb}[2]{
\vspace{0.05cm}
    \begin{tcolorbox}[
        left=2pt, right=2pt, top=2pt, bottom=2pt,
        boxrule=0.2mm,
        leftrule=2mm,
        arc=0mm,
        colframe=black!40!white, 
        colback=black!5!white, 
        colbacktitle=black!50!white 
    ]
    \textbf{#1}{#2}
    \end{tcolorbox}
\vspace{0.05cm}
}

\begin{document}

\title[\name: Partial Code Execution via Agentic Context Generation]{AgentExecutor: Partial Code Execution via\\Agentic Context Generation}
\author{Junkai Chen}
\email{junkaichen@smu.edu.sg}
\orcid{0009-0000-9945-7729}
\affiliation{%
  \institution{Singapore Management University}
  \country{Singapore}
}
\author{Chengran Yang}
\email{cryang@smu.edu.sg}
\affiliation{%
  \institution{Singapore Management University}
  \country{Singapore}
}
\author{Xing Hu}
\authornote{Corresponding author.}
\email{xinghu@zju.edu.cn}
\affiliation{%
  \institution{Zhejiang University}
  \country{China}
}
\author{Zhenhao Li}
\email{lzhenhao@yorku.ca}
\affiliation{%
  \institution{York University}
  \country{Canada}
}
\author{Xin Xia}
\email{xin.xia@acm.org}
\affiliation{%
  \institution{Zhejiang University}
  \country{China}
}
\author{David Lo}
\email{davidlo@smu.edu.sg}
\affiliation{%
  \institution{Singapore Management University}
  \country{Singapore}
}
\renewcommand{\shortauthors}{Chen et al.}

\begin{abstract} 
Executing code snippets is essential for dynamic program analysis, but it remains challenging to execute an arbitrary code snippet due to issues like missing context and incomplete dependencies. 
Existing approaches to partial code execution, such as LExecutor and Treefix, leverage the power of language models to infer missing information and enable execution. However, they suffer from (i) limited action spaces and feedback, and (ii) rigid optimization strategies, which restrict their effectiveness and efficiency.

In this paper, we propose~\name, a novel multi-agent framework for partial code execution. Our approach introduces a three-phase design: execution environment preparation, dynamic exploration with iterative refinement, and prefix evolution via program synthesis. 
Supported by the power of LLM agents who can think, act, and get feedback iteratively, \name~is able to autonomously explore a richer action space, enabling diverse operations such as creating resource files and resolving environment configuration. Furthermore, it adopts adaptive optimization strategies, including coverage-guided context pruning and prefix evolution via program synthesis, to systematically improve the execution quality of partial code. 

We evaluate~\name~on two widely used datasets comprising Stack Overflow snippets and open-source project code. The results show that~\name~achieves up to 94\% and 90\% code coverage, outperforming the state-of-the-art approach Treefix by 19.9\% and 13.8\%, respectively. In addition, \name~significantly reduces execution time (by up to 80.3\%) and cost (by up to 56.6\%). These findings demonstrate that \name~provides an effective and efficient solution for partial code execution.
\end{abstract}

\ccsdesc[300]{Software and its engineering~Software verification and validation}
\keywords{Partial Code Execution, Code Agent, Large Language Model}


\maketitle

\section{Introduction} 

Code execution provides access to the runtime information of the program, which supports a variety of dynamic analysis tasks, such as fuzz testing~\cite{godefroid2008automated,xia2024fuzz4all,meng2024large}, program slicing~\cite{agrawal1990dynamic,yadavally2024predictive,sintaha2023katana}, and performance analysis~\cite{graham1982gprof,gong2015jitprof,huang2021tprof}, facilitating software maintenance and development. For example, developers can use breakpoints to collect the program states for debugging.
However, if given an arbitrary code snippet, it is typically nontrivial to execute it due to issues like missing contexts or environmental factors.
When we search for answers to programming questions on Stack Overflow~\cite{stackoverflow}, 
these answers typically contain only illustrative solutions, i.e., code without implementation details or environment configuration, and it is typically infeasible to use them directly as they are. 
With regard to code snippets from larger-scale projects on GitHub, even though the full artifacts required for execution are provided, there could be issues that prevent their execution as well, such as dependency conflicts and environment incompatibilities. 
In these cases, extra manual effort is required to make these \textit{partial code snippets} executable, which is typically tedious and time-consuming.

To enable the automatic execution of arbitrary code snippets, various approaches have been proposed to support this task, i.e., 
partial code execution~\cite{souza2023lexecutor,souza2025treefix,groninger2025changeguard,xue2024selfpico,hayet2024feedback}. 
LExecutor~\cite{souza2023lexecutor} is a pioneering work applying learning-based approaches to this field. 
It trains neural language models to predict the values of variables under certain code contexts, and then utilizes them to inject their predictions into the runtime of the program to keep the execution proceeding. 
This research has inspired a line of subsequent research~\cite{souza2025treefix,groninger2025changeguard,xue2024selfpico,hayet2024feedback}. 
A recent work, Treefix~\cite{souza2025treefix}, leverages the power of large language models (LLMs) to create a prefix tree of code context and achieves very promising results. 
These works also explore applying their techniques to various downstream tasks, including bug reproduction~\cite{hayet2024feedback}, regression testing~\cite{groninger2025changeguard}, and runtime error detection~\cite{xue2024selfpico}, showcasing the practical application of partial code execution in real-world software engineering processes. 

While previous research provides valuable insights and initial solutions, we uncover two major limitations through an analysis of the state-of-the-art solution, i.e., Treefix~\cite{souza2025treefix}:

(i) \textit{\textbf{Limited action space and feedback}}. 
While Treefix leverages the strong capability of LLMs, its action space is restricted and fixed, which consequently leads to rather uninformative feedback and hinders potential improvements to the prefixes. 
For example, Treefix only allows two fixed types of contexts in its prefix (i.e., library import and variable initialization), ignoring other types of contexts. 
It also has no access to the running environment, which hinders many necessary actions to support code execution, like creating resource directories and searching certain code blocks. 
Such a limitation is also evident in aspects like handling dependency resolution and identifying undefined elements, which both rely solely on inflexible, static ways (e.g., using the static tool \texttt{pipreqs} as the only means to identify missing libraries).

(ii) \textit{\textbf{Rigid optimization strategy}}. 
Treefix optimizes code prefixes using a tree-based expansion, but it follows a largely rigid, fixed strategy. 
For instance, Treefix invokes the LLM a fixed number of times (e.g., 10), regardless of the prediction phase and current coverage status, which is wasteful when the optimization has already saturated, and potentially insufficient when further refinement is still needed.
Moreover, it depends solely on the generation from LLMs as the prefix (i.e., using LLMs to predict the prefix directly) without other synthesis strategies. This single approach makes it difficult to systematically traverse potentially valuable candidates and increase the quality of partial execution.

To address the aforementioned limitations, this paper presents\\ ~\name, an effective and efficient approach for partial execution supported by agentic coordination (where multiple agents are capable of thinking, acting with tools, and retrieving execution feedback in an iterative fashion).
\name~generates the context for partial code by searching for code prefixes with a three-phase design: (i) \textit{\textbf{execution environment preparation}}, where it prepares the environment by attempting to execute, (ii) \textit{\textbf{dynamic exploration with iterative refinement}}, where our method dynamically discovers diverse program paths for better prefixes, and (iii) \textit{\textbf{prefix evolution via program synthesis}}, where~\name~evolves prefixes by generating a prefix generator. 
Specifically, 
to mitigate the issue of \textit{limited action space and feedback}, with the aid of LLM agents and tool usage capability,~\name~can autonomously explore a diverse action space and get corresponding feedback for better prefixes, e.g., installing dependencies, creating auxiliary directories, and trying different solutions. 
Furthermore,~\name~addresses \textit{rigid optimization strategy} by adopting novel, dynamic optimization strategies for prefix refinement: 
In the phase of prefix exploration, it prunes visible contexts of agents guided by coverage improvement, preventing possible attention dilution of LLMs. 
In addition, in the final phase,~\name~systematically searches for better prefixes by synthesizing a prefix generator. We design a coverage-based method to select high-value prefixes as seeds for the generator to motivate better generations.

We evaluate~\name~on two datasets widely adopted in previous research~\cite{souza2023lexecutor,souza2025treefix}, consisting of partial code snippets drawn from popular open-source projects and Stack Overflow posts. 
Evaluation results show that \name~can cover 94\% and 90\% of lines of code within a single code prefix for Stack Overflow snippets and open-source functions, respectively.
Under the same performance metric, compared with the previous state-of-the-art approach Treefix~\cite{souza2025treefix},~\name~achieves relative improvements of 19.9\% and 13.8\%, while significantly reducing the average execution time (i.e., -80.3\% \& -77.9\%), cost (i.e., -56.6\% \& -52.1\%), and the numbers of LLM invocations (i.e., -52.1\% \& -53.9\%) on the two datasets. 
Moreover,~\name~can fully execute (i.e., achieving 100\% line coverage) 36.1\% and 18.1\% more partial code snippets on functions from the two datasets, respectively. 
These results demonstrate that~\name~presents substantially improved effectiveness on the task of partial code execution, while achieving improved efficiency on time and cost in comparison with baselines. 
We also provide analyses of our key design choices, accompanied by two case studies for the illustrative demonstration of~\name.

In summary, we make the following contributions:
\begin{itemize}[leftmargin=*]
    \item We propose~\name, a multi-agent approach for partial code execution, which has access to a wide action space and adopts well-designed optimization strategies.
    To the best of our knowledge, it is the first agentic approach for this task.
    \item Compared with the previous state of the art,~\name\\presents a significant improvement on code coverage (i.e., up to 20.2\%), while showing much better efficiency in time, cost, and LLM invocation times (i.e., up to 80.3\% reduction).
    \item To facilitate further research, we publicly open-source our code and datasets at GitHub and Zenodo~\cite{replication}.
\end{itemize}



\section{Related Work}

\subsection{Partial Code Execution}
The technique of partial code execution enables the execution of incomplete code snippets. 
LExecutor~\cite{souza2023lexecutor} proposed the idea of learning-guided execution that leveraged deep learning models to predict the runtime value of partial code. It then injects the prediction during the execution process to continue the program running. 
Hayet et al.~\cite{hayet2024feedback} proposed Incompleter to use execution feedback to guide the running of code. It consists of one mocker and one unmocker to enable execution while preserving the semantics of the original code. 
SelfPiCo~\cite{xue2024selfpico} applies large language models to partial code execution. In this work, an LLM is used as an interactive value predictor, which iteratively refines its prediction by an execution checker. 
ChangeGuard~\cite{groninger2025changeguard} extends the scope of partial code execution to a pair of code in the code change. By the technique of pairwise execution, which compares the dynamic behaviors of two paired programs, this work helps determine whether one code change is semantic-preserving or not. 

The latest work of partial code execution refers to Treefix~\cite{souza2025treefix}. 
This work constructs a tree of prefixes of the program to guide the code execution. During the expansion of the prefix tree, it tries to incrementally solve execution issues (e.g., variable undefinition) and improve the execution metrics (e.g., code coverage). 
Compared with Treefix,~\name~utilizes multiple LLM agents with wide action space and dynamic optimization strategies.

\begin{figure*}[t]
\centerline{\includegraphics[width=0.9\linewidth]{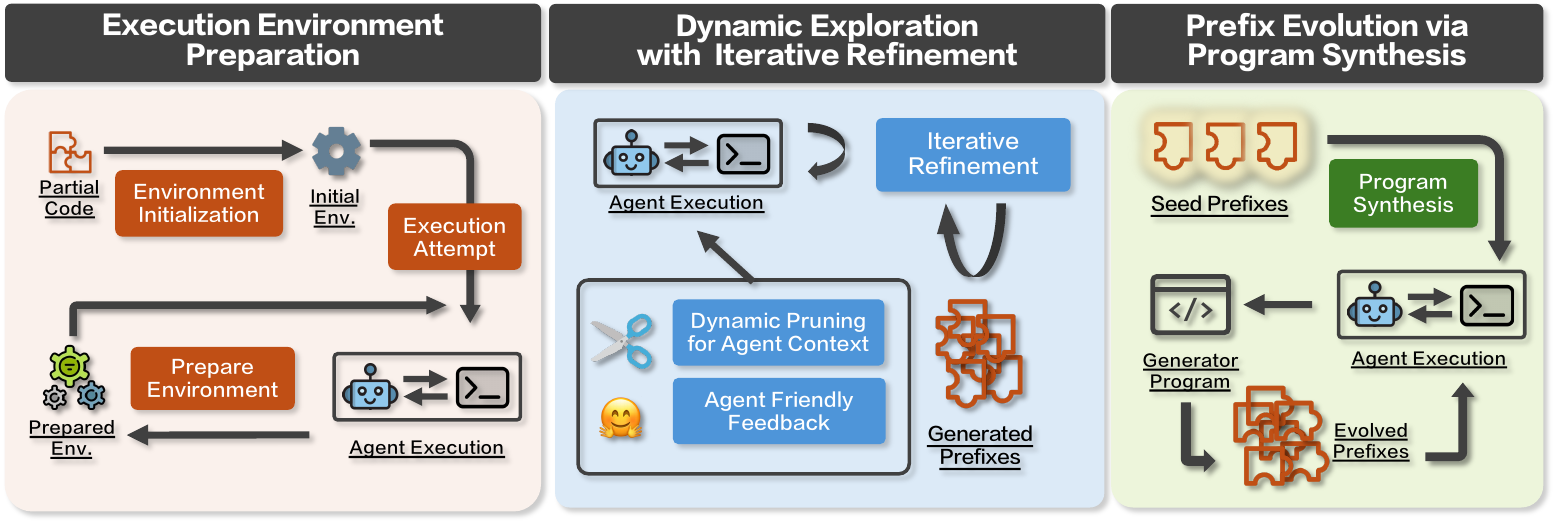}}
\caption{Overview of~\name, which consists of three phases: (i) Execution Environment Preparation, (ii) Dynamic Exploration with Iterative Refinement, and (iii) Prefix Evolution via Program Synthesis. 
Given the partial code, we first (i) prepare the execution environment by attempting to execute, supported by EnvAgent, then (ii) iteratively generate and refine prefixes where we design dynamic pruning and friendly feedback for RefAgent, and finally (iii) produce a generator program to evolve more prefixes from selected seed prefixes.
}
\label{figure:overall}
\end{figure*}

\subsection{Test Generation and Fuzzing} 
Fuzzing and test harness generation automatically produce inputs and driver code to dynamically test software. 
Generator-based fuzzers execute generator programs to create structured inputs~\cite{quickcheck,zest}, and recent work synthesizes such generators with LLMs~\cite{elfuzz}; harness generation constructs fuzz drivers for target APIs from client code, unit tests, or LLMs~\cite{fudge,fuzzgen,utopia,zhang2024fuzzdriver,promptfuzz}. Both assume a complete, executable target with well-defined entry points, and aim to expose bugs. In contrast, \textsc{AgentExecutor} targets incomplete snippets: it synthesizes a generator program to evolve prefixes that supply missing context, making arbitrary partial code executable and maximizing its own coverage.

\subsection{Agent-Based Approach for Software Engineering} 

Empowered by capabilities like tool use and planning, LLM-based agents have shown great capabilities in various software engineering tasks~\cite{hou2024large, hu2025assessing,liu2024large,chen2026securevibebench,chen2024code,chen2025reasoning,Tan2026LLMRCA,26TanLIDL}. 
RepairAgent~\cite{bouzenia2025repairagent} is an autonomous LLM-based agent framework for automated program repair. It formulates program repair as an agentic process where the model iteratively localizes bugs, gathers repair-relevant context, and validates candidate patches through tool use and execution feedback. 
Zhang et al. introduced AutoCodeRover~\cite{zhang2024autocoderover}, a code agent for resolving software engineering tasks such as fixing bugs in GitHub. The system combines LLM-based reasoning with structured code retrieval and spectrum-based fault localization to identify relevant program elements and generate patches in an iterative manner. 
VulTrial~\cite{widyasari2025let} introduces a multi-agent approach to simulate a court to review and examine the security issues in the program. The roles (e.g., code author) in the court are supported by agents who first discuss with each other, and then the final review board will give a final decision about whether one program is vulnerable or not. 
In this work, we apply agent-based approaches to the field of partial code execution.


\section{Approach}

\begin{figure}[t]\centerline{\includegraphics[width=0.8\linewidth]{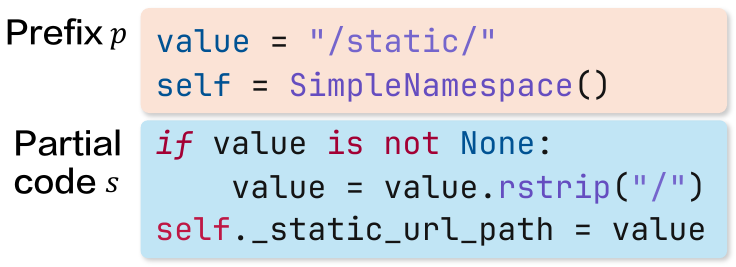}}\caption{An example of incomplete code \textit{s} and the expected prefix \textit{p}.}\label{figure:example}\end{figure}

\subsection{Problem Description} 
We first introduce the problem of partial code execution. 
Given an incomplete code snippet \textit{s}, this task aims to add a prefix \textit{p} (e.g., \texttt{import} statements to use libraries, value assignments to variables used in the code) to make the complete code \textit{p+s} executable, while maximizing the lines of code executed in the original part \textit{s}. 
For example, in Figure~\ref{figure:example}, the incomplete snippet
$s$ references two undefined variables, \texttt{value} and
\texttt{self}. The prefix $p$ resolves this by assigning
\texttt{value} a concrete string and instantiating \texttt{self} as a
\texttt{SimpleNamespace} object, so that $p+s$ executes successfully.
The coverage of the added prefix does not count, because the incomplete code \textit{s} is the target program we would like to execute and study, while the prefix serves as an auxiliary tool to achieve this goal.

Sometimes one isolated prefix \textit{p} cannot cover the full code lines in \textit{s}, e.g., there are two \texttt{if} branches with conditions that conflict with each other.
Therefore, consistent with the setting of Treefix~\cite{souza2025treefix}, we also generate a set \textit{P} of prefixes and count the union of covered lines for $p\in P$. We call this the \textit{Cumulative Coverage} (C-C); correspondingly, \textit{Single-Best Coverage} (S-C) refers to the best coverage reached with one single prefix.
At its core, for one incomplete code snippet,~\name~will produce a set of prefixes and optimize both its single-best and cumulative coverages. 

\subsection{Overview of~\name} 
\name~is a multi-agent approach for partial code execution, illustrated in Figure~\ref{figure:overall}.
Overall, it consists of three stages, each of which is responsible for different roles to execute the partial code and maximize the coverage. Firstly, we initialize and prepare the execution environment with the help of the Environment Agent (EnvAgent); then, a Refinement Agent (RefAgent) will iteratively generate and refine the prefixes; finally, we create an Evolution Agent (EvoAgent) to synthesize a generator program to produce more potential prefixes. 
Given the substantially different tasks handled by the three agents, we adopt a multi-agent rather than single-agent architecture, which allows for customized context management, reduces attention dilution and extra cost, and enables each agent to be powered by the model best suited to its task.

\name~has two distinct advantages over previous baselines: 
\begin{itemize}[leftmargin=*]
    \item \phead{Wide Action Space and Feedback.} 
    In Treefix, the action space is strictly limited for the execution environment (e.g., relying solely on static dependency resolution) and prefix refinement (e.g., allowing only two types of code statements). 
    ~\name~resolves this by using LLM agents with diverse commands in all phases. For example, it can create required resource files, view and edit programs, and install dependencies flexibly. We also analyze the used action space in Section~\ref{subsection:design} and introduce a case study in Section~\ref{subsec:case1}. 
    Besides, we design an agent-friendly feedback mechanism to make~\name~better understand the current failure and decide the next action (introduced in Section~\ref{subsubsec:iterative}). 
    \item \phead{Dynamic Optimization Strategies.} Treefix follows a rigid way to optimize its prefix (e.g., calling LLMs a fixed number of times regardless of current
    coverage improvement).~\name~proposes two novel strategies to achieve a flexible optimization: 
    (i) To avoid the attention dilution of LLM agents, we adopt a coverage-aware context pruning technique to dynamically crop the history of low value (introduced in Section~\ref{subsubsec:pruning}); 
    (ii) To achieve a systematic mining of generated prefixes,~\name~produces a prefix generator program to iterate all valuable prefixes without numerous LLM invocations (introduced in Section~\ref{subsubsec:generator}). 
    These two strategies show superior effectiveness while significantly reducing the time and monetary cost (which we discuss in Section~\ref{subsec:efficiency} and~\ref{subsec:effectiveness}).
\end{itemize}

\subsection{Execution Environment Preparation} 
\label{subsec:envagent}
This stage prepares the execution environment of the incomplete code for the subsequent actions. It is responsible for the environment management of code execution, including initializing
virtual environments, installing required libraries, and solving environmental issues like missing files. 
It prepares the environment by
iteratively attempting to execute code and solving issues based on
execution feedback.

\phead{Environment Initialization.} 
Before code execution and analysis, we first need to set up the environment for~\name~ to run the program and obtain execution feedback. 
In Treefix, they use a shared environment between different snippets; although it can accelerate the dependency installation and avoid repeated actions, the environment sharing may bring certain issues like contaminating other workspaces and dependency conflicts.
To avoid these environmental issues, we build an isolated virtual environment for each code snippet. 
Specifically, we prepare a cleaned working directory and use \textit{uv}~\cite{uv_astral_2026}, a popular environment manager, as the environment backbone to manage the code execution environment, such as package installation and version switch. 
Compared with other solutions such as \textit{venv}~\cite{python_venv} and \textit{conda}~\cite{conda}, uv enables extremely fast package management at the project level, reducing environment setup time in our use case.
We also pre-install common-use packages like 
\texttt{numpy} to reduce the latency in subsequent agent actions.

\phead{Preliminary Execution Attempt.} 
After the environment initialization,~\name~will attempt to execute the code with EnvAgent.
EnvAgent will first inspect the working directory and the partial code in order to understand what is missing in the environment for executing code, then it will try to take actions to manage this execution environment. 
The agent realizes these through the bash tool, where it can execute commands (e.g., \texttt{pip install} in the terminal) and obtain execution feedback (e.g., a \texttt{NameError}). 
Typical operations in the environment could include package installation and file creation; however, due to the wide action space with bash, it is able to do a lot of things, leading to the environment preparation. We also explore this in Section~\ref{subsection:design}.
%
After this, EnvAgent will try to run the code in the virtual environment directly. 
If the agent thinks the code runs successfully and no additional actions are required, it will finish this phase on its own. 
Otherwise, based on execution feedback, EnvAgent will consider the potential cause and propose an action to address it until success or the step limit is reached.

\begin{figure}[t]
\centerline{\includegraphics[width=0.8\linewidth]{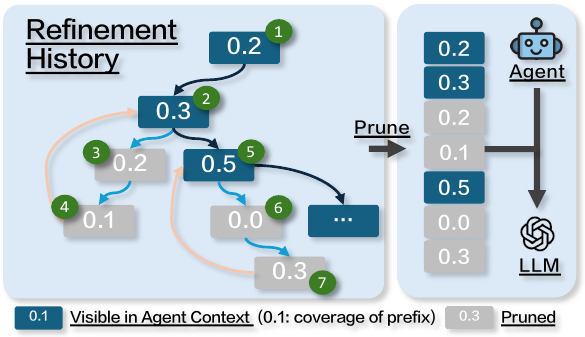}}
\caption{Illustration of dynamic pruning. ``\ding{172}\ding{173}...'': numbers represent the order of exploration.}
\label{figure:pruning}
\end{figure}

\subsection{Dynamic Exploration with Iterative Refinement} 
\label{subsec:refagent}
With the execution environment prepared using EnvAgent, this phase primarily focuses on exploring different program paths and generating various prefixes using agent-based iterative refinement.

\subsubsection{Iterative Refinement with Agent Friendly Feedback.} 
\label{subsubsec:iterative}
In this step, RefAgent reasons about the program paths and current situations, generates different prefixes, tests their coverage, and retrieves the execution feedback for further refinement. 
Specifically, we prompt RefAgent to inspect the partial code and think about potential program execution paths, especially key elements that are not initialized or not properly used (e.g., the condition variable in the \texttt{if}). 
After that, based on this thought about program paths and the previous history if available (e.g., last turn of action and feedback), RefAgent attempts to generate or refine the code prefix via action execution (e.g., use \texttt{sed} to edit the file), and then retrieve the corresponding feedback. 
If the coverage is successfully returned and satisfactory to RefAgent, it will finish its task; otherwise, it appends the thoughts, actions, and feedback to the context history, proceeding to the next refinement iteration.

In RefAgent, the feedback is an important factor for refining the prefix. The feedback can include the execution status of code, an illustration of code coverage, general execution traces from bash commands, etc. 
For instance, when the coverage is not full, and there are still lines not covered, we follow Treefix~\cite{souza2025treefix} and use special marks in the comments to label the corresponding lines of code in order to provide targeted feedback. 
However, when the agent executes commands directly in the terminal, the returned feedback (e.g., a long trace stack) is typically hard to read and understand; therefore, we design several agent-friendly feedback mechanisms for RefAgent to better understand the current situation and prepare the next actions. 
For example, we follow Treefix~\cite{souza2025treefix} and use static analysis to identify undefined variables in the partial code; if any one is undefined in the prefix, the environment will not only print out the raw outputs (e.g., throwing the \texttt{NameError}), but also prompt RefAgent to notice this issue with a tip about the potential inaccuracy for this warning (as the results of static analysis may not be accurate). 
When the output of the program is empty, RefAgent will receive the note to check if the code is executed successfully or if there are issues like silent early exits or missing exception handling.
We also design a specialized feedback if the generated prefix is not valid in format (e.g., invalid editing of partial code), leading to unexpected coverage computation failure. These small but useful designs significantly improve the instruction following ability of the LLM agents.

\subsubsection{Dynamic Pruning for Agent Context} 
\label{subsubsec:pruning}
During the iterative refinement with RefAgent, history messages are available in the context of the LLM. 
These historical messages include thoughts, actions, and feedback, and they serve as useful experiences to better process the current refinement or other actions.
However, if the history is too long and redundant, these contents will, in turn, bring negative effects to RefAgent (e.g., stuck in the error loop) and increase the consumption of time and monetary cost as well. 

Therefore, we propose to apply a coverage-aware dynamic pruning for the history context in RefAgent.
As shown in Figure~\ref{figure:pruning}, the agent first explores one prefix refinement direction and obtains their coverage results (i.e., the most left path of "\ding{172}\ding{173}\ding{174}\ding{175}"). 
Then, once newly explored prefixes are worse than the previous best for several consecutive attempts (e.g., 0.1 and 0.2 < 0.3, with patience = 2 in this illustration), motivated by the idea of \textit{early stopping}, we will prune the two failed exploratory messages and take previous successful, coverage-increasing history as the new context for refinement (i.e., "\ding{172}\ding{173}"). Similarly, the second pruning point is "0.3" on the path "\ding{172}\ding{173}\ding{176}\ding{177}\ding{178}" as RefAgent has no patience to wait for a better prefix with coverage over the previous best ``0.5". 
For the next exploration, RefAgent has the visible context ``\ding{172}\ding{173}\ding{176}''.
This mechanism helps mitigate the issue of attention dilution for LLM agents while helping reduce the extra cost brought by a longer context. 
We set a limited budget (i.e., patience) for the agent to explore with the context of the unsuccessful history, which is considered a trade-off between free exploration and guided improvement.

\subsection{Prefix Evolution via Program Synthesis} 
\label{subsec:evoagent}
In this phase, based on existing ones generated by RefAgent, we use EvoAgent to evolve more prefixes with the help of program synthesis.  

\subsubsection{Coverage-Aware Prefix Seed Selection} 
\label{subsubsec:seed}
%
We first construct a set of prefixes as the seeds for the evolutionary process. 
Intuitively, the seed set should be sufficiently representative to guide the agent toward generating more effective prefixes, while remaining compact to avoid introducing too much context in the program synthesis.
During the exploration stage, RefAgent generates a large number of prefix candidates. However, directly using all these candidates as seeds is undesirable because some of them could be redundant or contain noisy behaviors, and including all of them may provide little additional benefit. Therefore, we select only a small subset of high-value prefixes to initialize the evolutionary process.

Intuitively, prefixes that trigger the execution of more distinct code lines are more likely to represent different program behaviors. 
Our key intuition is that a good seed set should expose the agent to diverse behaviors of the target program. 
To select these seeds, we approximate behavioral diversity using code coverage. 
If the seeds collectively cover more execution behaviors, the evolutionary search is more likely to discover new and effective prefixes. 
Therefore, we aim to select a small set of prefixes whose combined code coverage is maximized. At the same time, the number of selected prefixes should remain limited so that the prompt context stays concise and manageable for the agent.
Specifically, we view each prefix as the set of lines it covers; given a budget of at most N seeds, the goal is to maximize the size of their union. To improve efficiency, we use dynamic planning to implement this strategy.

\subsubsection{Evolution Program Generation.} 
\label{subsubsec:generator}
After several representative prefix seeds are filtered out, we apply EvoAgent to evolve more prefixes based on these prefixes via program generation and execution. There are three main steps: (i) analyze key elements and their values, (ii) produce the generator program to iterate all combinations, and (iii) verify and refine the generator program. 

\begin{figure}[]
\centerline{\includegraphics[width=0.8\linewidth]{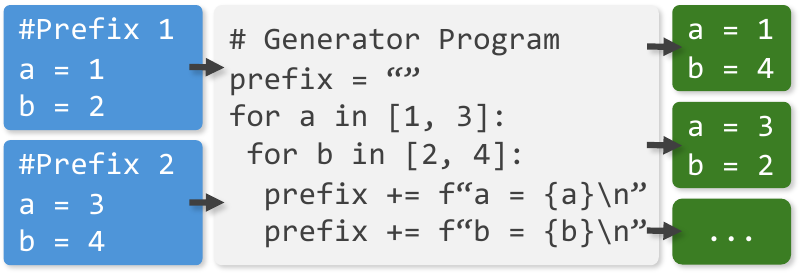}}
\caption{A simplified, illustrative example of the prefix evolution.}
\label{figure:evolution}
\end{figure}

\phead{Analyze Key Elements.} 
First, EvoAgent understands key elements in the partial code. We explicitly prompt the agent to reason about key variables related to the program execution path (e.g., a variable in one control flow) and their potential values given selected prefix seeds, and let EvoAgent attempt to predict all possible occurrences of these variables (e.g., their assigned values). 
Through this step, we aim to ensure that EvoAgent comprehends diverse program paths in the partial code and analyzes potential reachability with prefixes. 

\phead{Produce Program for Evolution.}
After that, EvoAgent will produce a generator program to help implement all possible combinations via executing the generator program. Figure~\ref{figure:evolution} provides a simplified, illustrative example of this process. 
There are two prefix seeds: \texttt{a = 1, b = 2} and \texttt{a = 3, b = 4} shown in the left part of the figure. 
EvoAgent will first identify the variables in the prefix seeds (i.e., \texttt{a} and \texttt{b}) and their potential values (i.e., \texttt{a} has two possible values: \texttt{\{1, 3\}}, and {b} is for \texttt{\{2, 4\}}), and then it will generate a Python script to iterate all these combinations. In this example, EvoAgent uses two \texttt{for} nested loops to generate all potential combinations, leading to two new prefix candidates: \texttt{a = 1, b = 4} and \texttt{a = 3, b = 2}. 
In practice, this process is typically more complicated and diverse, involving various types of identified elements, assigned values, implementation of the generator, etc. As the LLM agent has full autonomy to realize different generator programs and retrieve the coverage feedback, it helps~\name~conduct a systematic search to synthesize all potential prefix candidates; compared with previous approaches,~\name~innovatively utilizes the program synthesis ability of LLM agents to produce a prefix evolution generator, which extends the search space and reduces the cost compared with directly obtaining predictions from LLMs.

\phead{Verify and Refine the Generator.} 
For one generator program, EvoAgent will try to execute it, inspect its generated prefixes, and use the provided coverage tool to check their coverage. If the execution of the generator fails, or EvoAgent thinks that the resulting coverage of generated prefixes is not satisfactory, it will iteratively refine the generator program, execute and obtain generated prefixes, and test their coverage, until the agent believes there is little room for further improvements.

Finally,~\name~will automatically save all the generated prefixes EvoAgent produces and then re-calculate the coverage of them to find the best covered code prefix. 
As the number of prefixes could be large, we adapt the original coverage computation script provided in previous works~\cite{souza2025treefix} with the help of multi-process execution. 

\subsection{Coverage Assessment and Agent Scaffold} 
\phead{Coverage.}
For each code snippet and a set of prefixes generated during the full pipeline, we will collect all the generations and compute the coverage again. 
This step performs a secondary confirmation of results, avoiding acquiring wrong results caused by potential invalid operations from LLM agents.
We compute the line coverage in the scope of the original program (i.e., not including the surrounding prefix and suffix), consistent with the evaluation setting in Treefix~\cite{souza2025treefix} and LExecutor~\cite{souza2023lexecutor}. 
Specifically, the program is instrumented by a special function \texttt{\_l\_()} after every code line. This function will be called if one line of code is successfully executed. 
Following the same setup, we note that this setup is stricter than general line coverage computation, as we only consider that one line is covered if the code line is executed without crashing. 
In our evaluation, we compute the code coverage under one execution and multiple attempts; please check Section~\ref{setup:metrics} for more details about the metrics.

\phead{Agent Scaffold.}
\name~adopts multiple LLM agents to enable partial execution in different roles.
These agents share the same base scaffold, while we design specialized harnesses (e.g., prompt design, context management, tool use) separately to let them fit in different responsibilities.
Specifically, for the selection of agent scaffold, we adopt a ReAct~\cite {yao2022react}-like framework mini-SWE-agent~\cite{yang2024sweagent} as the base agent, which is popular among the software engineering community and presents powerful capability~\cite{jimenezswe2024}. 
We use this framework because it offers a minimal implementation of a basic agent framework (i.e., execution in a simple loop of thinking and acting with the bash tool), providing easy access for adaptation and not relying on any particular design. Besides, we include all original configurations, including prompts, in our replication package.







\section{Experimental Setup}

\subsection{Baselines} 

We compare~\name~with the following approaches\\adopted by previous works~\cite{souza2023lexecutor,xue2024selfpico,souza2025treefix} based on our literature review in the field of partial code execution. 
We use the same experimental settings of these baselines adopted in the work Treefix~\cite{souza2025treefix} unless otherwise specified. 
\begin{itemize}[leftmargin=*]
    \item \textbf{Treefix}~\cite{souza2025treefix} is the state-of-the-art approach for partial code execution. 
    It designs a three-stage pipeline to identify missing elements in the code and iteratively prompt the LLM to generate code prefixes based on execution feedback. 
    \item \textbf{SelfPiCO}~\cite{xue2024selfpico} dynamically guides partial execution supported by LLMs within an iterative loop. It also uses in-context learning and chain-of-thought reasoning to inject human knowledge and enhance the reasoning capability. 
    \item \textbf{Incompleter}~\cite{hayet2024feedback} defines a series of rules (e.g., add class and file) to handle similar error patterns during execution. It then designs a feedback-driven approach from these error rules. 
    \item \textbf{LExecutor}~\cite{souza2023lexecutor} proposed the idea of learning-guided execution, i.e., using a learning-based approach to executing arbitrary code snippets. LExecutor trains language models to predict the missing values of partial code (e.g., undefined variables) and inject them to proceed with the execution.
    \item \textbf{Type4Py}~\cite{mir2022type4py} is a neural model-based approach for type inference. It adopts a hierarchical neural network to learn the high-dimensional features of types for arguments, variables, and return values. 
    \item \textbf{Pynguin}~\cite{lukasczyk2020automated} is a test generation tool for Python functions. It adopts a search-based algorithm and leverages type information to maximize the test coverage.
    \item \textbf{``As Is''} means we try to execute the original code without any modification, just as it is. 
    We can successfully run code directly if the code does not miss the required elements for execution. 
\end{itemize}

\subsection{Datasets}

\begin{table}[t]
\centering
\setlength{\tabcolsep}{4pt}
\caption{Datasets used for evaluation.}
\label{table:datasets}
\resizebox{\columnwidth}{!}{%
\tabcolsep=12pt
\begin{tabular}{lrr}
\toprule
\textbf{Dataset} & \textbf{\# of Snippets} & \textbf{\# of Avg. LoC} \\
\midrule
Open-Source Functions   & 1,000 & 9.7 \\
Stack Overflow Snippets & 462   & 7.5 \\
\midrule
Total                   & 1,462 & 9.0 \\
\bottomrule
\end{tabular}
}
\end{table}

We follow Treefix~\cite{souza2025treefix} and conduct experiments on the two datasets with the same dataset processing method. Statistics are presented in Table~\ref{table:datasets}.

\begin{itemize}[leftmargin=*]
    \item \textbf{Stack Overflow snippets dataset} includes 462 code snippets from the answers of Python programming questions collected from Stack Overflow~\cite{stackoverflow}.
    \item \textbf{Open-source functions dataset} contains 1,000 functions sampled from five open-source Python projects (i.e., \textit{black, tensorflow, scrapy, flask, and pandas}) on GitHub. 
\end{itemize}

\subsection{Metrics}
\label{setup:metrics}
Consistent with the work Treefix~\cite{souza2025treefix}, we adopt three coverage-based metrics to assess our approach in the main experiments.
\begin{itemize}[leftmargin=*]
    \item \textbf{Single-Best Coverage} (\textit{S-C}): For each code snippet, we compute the coverage of all predictions (i.e., code prefixes) and take the prediction with the highest coverage as its S-C result. This metric reflects scenarios focused on a single execution, e.g., inspecting one particular program path that requires only a single input. We then report the average S-C score over all code snippets in the dataset. S-C is consistent with ``the coverage of single-best prefix'' in the Treefix paper.
    \item \textbf{Cumulative Coverage} (\textit{C-C}): As we have a set of prefix predictions for one code snippet, C-C considers one code line to be covered if any prefix in these predictions covers it (i.e., the union). 
    Unlike S-C, which reflects a single execution trace, C-C is applicable in tasks where multiple inputs are jointly used to achieve broader coverage (e.g., fuzzing). Again, this metric is the same as the ``cumulative coverage'' of all prefixes in Treefix.
    \item \textbf{Full Execution Rate} (\textit{FER}): If one code snippet achieves a 1.0 C-C, i.e., full line coverage with all prefixes, we deem it a full execution. We report the percentage of code snippets in the dataset that achieve full execution. FER is taken as a supplement to C-C. 
\end{itemize}

For all the metrics mentioned above, a larger value indicates better performance.

\subsection{Implementation Details}
\phead{Agent and LLM.} 
For each agent in~\name, we set the step limit to 70, consistent with Treefix~\cite{souza2025treefix}, which allows up to 210 (=70 times/agent * 3 agents) LLM responses.
For RefAgent, the patience for dynamic pruning is set to 10 based on a small set of preliminary experiments.
For backbone LLMs in~\name, we use GPT-5 nano \textit{(gpt-5-nano-2025-08-07)} for EnvAgent and 
GPT-5 mini \textit{(gpt-5-mini-2025-08-07)}~\cite{openai_gpt5mini} for the other two agents. 
Compared with other models like GPT-4o, these models have been optimized for agentic tasks, making them suitable for our usage scenario in this work. 
We choose to combine different model versions to achieve a balance between effectiveness and efficiency of~\name. 
For the reasoning effort, we set it to ``minimal'' to reduce token consumption and use default values for the other parameters.
We use OpenRouter~\cite{openrouter2026} and litellm~\cite{berriai_litellm_2024} to obtain unified and convenient access to endpoints of LLM servers, compatible and consistent with the agent scaffold. 

\phead{Environment.} 
We conduct our experiments in a Linux server within Dockerized environments. 
For experiments of Treefix and\\\name, to mitigate potential issues such as network instability, we prepare the cache of commonly used PyPI~\cite{pypi} packages and allow them to share it, which could help achieve a fair evaluation. 

\phead{Baseline.} 
The baseline Treefix originally uses GPT-4o series as the backbone LLMs, but this model family is kind of outdated; therefore, we rerun its experiments with the same backbone LLM, GPT-5 mini, and accordingly increase the context window Treefix can utilize to 128,000.
We keep all other configurations the same as it is to achieve a fair comparison.

\section{Results}

\begin{table*}[t]
\centering
\setlength{\tabcolsep}{4pt}
\caption{Overall effectiveness of our approach and baselines. 
For Treefix, we replicate the experiments with the same backbone model to conduct a fair comparison; for other baselines, we take results from this paper~\cite{souza2025treefix} under the same settings and omit inapplicable ones with the mark ``-''. We report the \textbf{relative improvement} of our approach with \underline{the second-best results} (with underline) for the corresponding columns. 
``S-C'': Single-Best Coverage; ``C-C'': Cumulative Coverage; ``FER'': Full Execution Rate. 
}
\label{table:overall}
\resizebox{0.95\linewidth}{!}{%
\tabcolsep=12pt
\begin{tabular}{@{}l|lll|lll@{}}
\toprule
 & \multicolumn{3}{c|}{\textbf{Stack Overflow Snippets}} & \multicolumn{3}{c}{\textbf{Open-Source Functions}} \\ \cmidrule(l){2-7} 
 & \textbf{S-C $\uparrow$} & \textbf{C-C $\uparrow$} & \textbf{FER $\uparrow$} & \textbf{S-C $\uparrow$} & \textbf{C-C $\uparrow$} & \textbf{FER $\uparrow$} \\ \midrule
As Is & 0.43& —& 0.30& 0.04& —& 0.02\\
Pynguin & —& —& —& 0.04& —& 0.02\\
Type4Py & 0.46& —& 0.32& 0.13& —& 0.08\\
LExecutor & 0.65& —& 0.49& 0.51& —& 0.35\\
Incompleter & 0.69& —& 0.53& 0.51& —& 0.35\\
SelfPiCO & 0.75& —& 0.60& 0.59& —& 0.40\\
Treefix & \underline{0.79} & \underline{0.79} & \underline{0.67} & \underline{0.79} & \underline{0.86} & \underline{0.68}\\
\midrule
\textbf{\name~(Ours)} & \textbf{0.94 (+19.9\%)}& \textbf{0.95 (+20.2\%)}& \textbf{0.91 (+36.1\%)} & \textbf{0.90 (+13.8\%)} & \textbf{0.93(+8.2\%)} & \textbf{0.80 (+18.1\%)}\\ 
\bottomrule
\end{tabular}%
}
\end{table*}
In this section, we propose and answer the following research questions (RQs):
\begin{itemize}[leftmargin=*]
    \item \textbf{RQ1}: What is the overall effectiveness of~\name~and baselines on covering partial code?
    \item \textbf{RQ2}: How efficient are~\name~and Treefix in terms of time consumption, monetary cost, and number of LLM invocations?
    \item \textbf{RQ3}: How do key design choices (i.e., wide action space provided, dynamic pruning, and prefix evolution via program synthesis) perform in~\name?
\end{itemize}

\subsection{RQ1: Overall Effectiveness} 
\label{subsec:effectiveness}
In this research question, we focus on the performance of baselines and~\name. Table~\ref{table:overall} shows the overall effectiveness results of~\name~and other baselines in terms of three code coverage metrics (i.e., S-C, C-C, and FER). We underline the second-best results of each column and report the relative improvements of our approach in comparison with them. 
We do not include results of some LLM-based approaches like Incompleter~\cite{hayet2024feedback} because they heavily rely on specialized, finetuned models and do not support API-based backbone language models.

Overall, ~\name~achieves the best performance across all three metrics on two different datasets with a large improvement. 
With regard to Single-Best Coverage (S-C) and Cumulative Coverage (C-C), results show that more than 90\% lines of code are covered by our approach, meaning that the prefixes~\name~generates are able to make most parts of the program executable in one or multiple execution attempts for both illustrative programs from Stack Overflow and functions in open-source projects. 
Besides, on the metric of Full Execution Rate (FER),~\name~also demonstrates good performance, with a 91\% for Stack Overflow snippets and an 80\% for open-source functions, respectively. 

Compared to the previous best technique, Treefix,~\name~has shown impressive improvement on all combinations of metrics and datasets. Shown in the brackets of the table, relative improvements of~\name~over Treefix range from 8.2\% (i.e., C-C in open-source functions) to 36.1\% (i.e., FER in Stack Overflow snippets), and the average number is 19.4\%. 
Taking into account that Treefix has outperformed the other two competitive methods (i.e., SelfPiCO~\cite{xue2024selfpico} and Incompleter~\cite{hayet2024feedback}) by a significant margin, we believe that the performance of~\name~is very impressive and promising.

\greyboxb{Summary of RQ1:~}{~\name~ presents new state-of-the-art results compared with Treefix in terms of code coverage with an up to 36.1\% relative improvement. It can cover more than 90\% lines of code on both Stack Overflow and open-source datasets.}

\subsection{RQ2: Efficiency Analysis}
\label{subsec:efficiency}
\begin{table*}[]
\centering
\setlength{\tabcolsep}{3pt}
\caption{Efficiency results of Treefix and~\name.}
\label{table:efficiency}
\resizebox{0.9\linewidth}{!}{%
\tabcolsep=12pt
\begin{tabular}{@{}l|lll|lll@{}}
\toprule
 & \multicolumn{3}{c|}{\textbf{Open-Source Functions}} & \multicolumn{3}{c}{\textbf{Stack Overflow Snippets}} \\ \midrule
 & \textbf{Time (s)} & \textbf{Cost ($10^{-2}$\$)} & \textbf{\# Call} & \textbf{Time (s)} & \textbf{Cost ($10^{-2}$\$)} & \textbf{\# Call} \\ \midrule
Treefix & 1124.01 & 12.60 & 75.42 & 704.60 & 4.86 & 45.84 \\ 
\textbf{AgentExecutor (Ours)} & \textbf{221.64 (-80.3\%)} & \textbf{5.48 (-56.6\%)} & \textbf{36.16 (-52.1\%)} & \textbf{155.56 (-77.9\%)} & \textbf{2.33 (-52.1\%)} & \textbf{21.13 (-53.9\%)}\\ \bottomrule
\end{tabular}%
}
\end{table*}
In this research question, we investigate the efficiency of \name~and the state-of-the-art baseline, Treefix, focusing on three key indicators: execution time (of the approach), monetary cost of API, and the number of LLM invocations. 
Other baselines like Type4Py~\cite{mir2022type4py} and LExecutor~\cite{souza2023lexecutor} are not included in this analysis because we think the large performance gap and differences in underlying techniques (e.g., heuristics, finetuning) make them somewhat incomparable.

We collect this information at the granularity of the task, which may consist of predicting and executing many candidate snippets for one partial code snippet. For example, ``execution time'' refers to the full execution time one approach consumes to give all predictions and find the best prefix. 
Note that this setting is different from that of Treefix~\cite{souza2025treefix}, where it computes the average among all execution attempts for one incomplete code snippet. For example, for one task corresponding to a specific incomplete code snippet, if the approach predicts and tests 10 candidates in 10 seconds, it will report the average time as 1 second (10 seconds/10 candidates). 
We believe our setup is more reasonable as it computes the consumption that the approach requires to achieve its best performance, and includes the potential loss for predictions that are invalid or without additional benefits.

Table~\ref{table:efficiency} presents the detailed statistics of the efficiency of Treefix and our~\name~with the relative changes. We introduce the corresponding results separately in the following parts:

\phead{Time Consumption.} 
For both~\name~and Treefix, we record the timestamps immediately before the processing of each incomplete code begins and after it finishes, and then compute their difference as the time consumption. 
In this process, the time for library installation, virtual environment creation (for each code snippet in~\name), and coverage computation is included as well. 

The time consumptions of~\name~are 221.64 seconds (s) and 155.56 seconds for open-source functions and Stack Overflow snippets, respectively. 
In comparison, Treefix requires around four to five times as long as~\name~(e.g., 221.64s vs. 1124.01s), and its average time consumption among the two datasets is more than 15 minutes (i.e., 900s). 

We attribute the unsatisfactory time efficiency of Treefix to the inflexible processing of the predictions. 
For one incomplete code snippet, Treefix applies the same strategy to process and test every predicted prefix: detect and install dependencies, execute and capture errors, filter incorrect lines of code, and measure the coverage. 
Considering the large search space of Treefix, this method is very time-consuming because many actions are not required to be repeated for the same original code. 
In our manual inspection of the runtime logs, we find that repeatedly detecting and installing the same dependencies—even when none are required—takes a significant amount of time during its execution. 
Treefix lacks a kind of feedback mechanism about this kind of information (e.g., whether the dependency is already installed). 
In comparison, even if our approach does extra work like creating isolated virtual environments,~\name~adopts a dynamic, agentic approach to achieve an efficient iterative loop while remaining effective.

\phead{Monetary Cost.} 
For~\name, we use the official interface provided by the agent scaffold to compute the cost for LLM tokens; 
For Treefix, we calculate this using the cost information returned by the LLM API. 
We choose not to use the cost computation script in the official repository of Treefix because it does not consider the varying price of prompt caching from the model provider (\textit{which is generally lower}), leading to an unfair comparison. 

To execute one partial code snippet and maximize the coverage,~\name~requires around 0.055 USD for open-source functions and 0.023 for Stack Overflow snippets, respectively. 
Compared with the results of Treefix (i.e., 0.0486 vs. 0.126 USD), the monetary costs of~\name~are less than half of them. 
We summarize two key designs of~\name~contributing to this: (i)~\name~dynamically prunes the redundant contexts during its iterative refinement based on coverage improvement, reducing the tokens used in the history of the agent while remaining performant; (ii) Instead of relying solely on direct generations from LLMs, EvoAgent in~\name~uses its code generation ability to efficiently produce a number of prefix candidates.

\phead{Number of LLM Invocations.} 
We report both token costs and numbers of LLM calls because they capture complementary aspects here: token consumption reflects the monetary cost of using LLMs, while the number of calls reflects the interaction overhead and potential latency introduced by repeated model invocations.

As shown in Table~\ref{table:efficiency},~\name~invokes LLMs 36.16 and 21.13 times for functions from open-source projects and Stack Overflow, respectively. 
Under the same request limits (i.e., 210 times in total), Treefix needs roughly double the calls of LLMs to finish the full process of one incomplete code snippet. 
The gap between~\name~and Treefix mainly comes from the design of prefix optimization: Treefix follows a fixed, tree-based expansion strategy for prefix refinement (i.e., 10 times for one prefix node), while~\name~utilizes LLM agents to dynamically and iteratively improve the prediction step by step, making full use of each feedback from LLMs.

\greyboxb{Summary of RQ2:}{
Compared with the baseline,~\name~consumes only around 20\% of the time and achieves a better performance. In the meantime, it also reduces more than half of the token costs and the number of LLM invocations, demonstrating its superior efficiency.
}

\subsection{RQ3: Design Choices}
\label{subsection:design}

\begin{table*}[t]
\centering
\setlength{\tabcolsep}{4pt}
\caption{Command Distribution of Different Agents.}
\label{table:commands}
\resizebox{0.85\textwidth}{!}{%
\tabcolsep=12pt
\begin{tabular}{lrrrrrr}
\toprule
\textbf{Agent} & \textbf{Execution} & \textbf{Text Processing} & \textbf{File I/O} & \textbf{File System} & \textbf{Statistics} & \textbf{Others} \\
\midrule
EnvAgent     & 6,779  & 5,489 & 3,291  & 1,912 & 2,560 & 354 \\
RefAgent & 18,520 & 687  & 34,797 & 2,322 & 53   & 7   \\
EvoAgent      & 7,929  & 1,769 & 5,070  & 672  & 101  & 11  \\
\bottomrule
\end{tabular}
}
\end{table*}
\phead{Wide Action Space with LLM Agents.} 
Table~\ref{table:commands} presents the actions (commands) that three agents used in the datasets. We collect all the commands and classify them into six categories, including execution (e.g., \texttt{python}, \texttt{bash}), text processing (e.g., \texttt{sed, awk}), File I/O (e.g., \texttt{cat, echo}), statistics (e.g., \texttt{wc, nl}), file system (e.g., \texttt{ls, cp}), and others (e.g., \texttt{curl, pip}). 
From the results, we find that: 
(i) The actions for agents are very diverse. These commands involve various steps~\name~required to test the code (execution), edit the code (text processing), create directories (file system), etc.
(ii) Three agents present different distributions of command usage. For example, ``Others'' exists much more frequently in EnvAgent than other agents because it requires the \texttt{pip} command to install dependencies; File I/O presents a significant proportion in RefAgent as it iteratively refines the prefix based on the execution feedback.

\begin{figure}[t]
\centerline{
\includegraphics[width=0.65\linewidth]{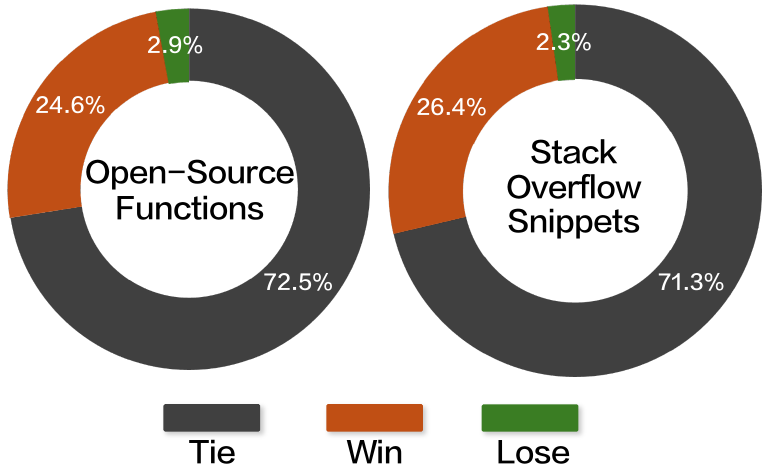}}
\caption{Distribution on whether pruning improves the following prefixes.}
\label{figure:111}
\end{figure}

\phead{Dynamic Pruning in Iterative Refinement.} 
We apply a dynamic pruning mechanism to reduce the issue of attention dilution while keeping~\name~efficient. To explore how this mechanism performs, we collect statistics about whether the future coverage after pruning can be better than the current best one, and results are shown in Figure~\ref{figure:111}. 
As illustrated, we find that in most cases, the future coverage could be tied with (i.e., around 72\%) or better than (i.e., around 25\%) the current status. 
It suggests that the mechanism of dynamic pruning can help reduce the time and cost, while keeping or increasing the coverage performance. 
We also find that there is less than 3\% of cases present a negative result. However, compared with neutral (i.e., ``Tie'') and positive (i.e., ``Win'') ones, its proportion could be negligible. 
Therefore, we think dynamic pruning brings essential benefits to~\name~on both effectiveness and efficiency.

\begin{figure}[t]
\centerline{\includegraphics[width=0.8\linewidth]{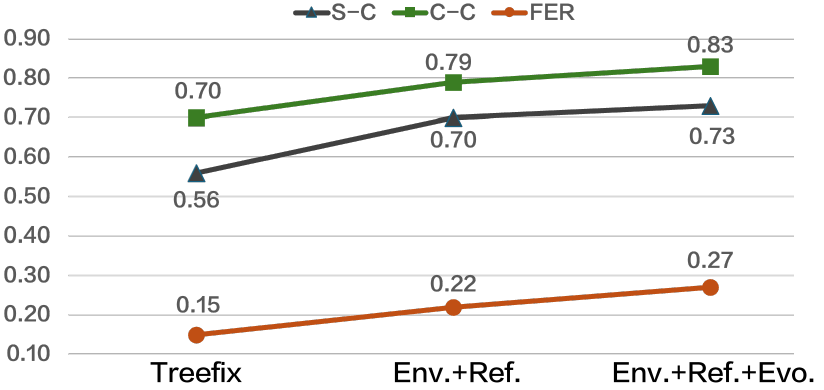}}
\caption{Results of Treefix,~\name~w/o EvoAgent, and the full AgentExecutor, on the subset of cases entering the evolution phase.}
\label{figure:ablation}
\end{figure}

\phead{Ablation Study for Prefix Evolution.}
We explore if the phase of prefix evolution takes effects to the coverage. 
For all data items in the two datasets, EvoAgent will not launch if the coverage has already achieved 1.0, and we find that there are around 14.1\% of them go into the third stage and use program synthesis to evolve more prefixes. 
For these cases, we collect their performance on Treefix,~\name~w/o EvoAgent (i.e., ``Env.+Ref.''), and the full~\name~(i.e., ``Env.+Ref.+Evo.''), and illustrate them in Figure~\ref{figure:ablation}. 
We have the following summaries: (i)~\name~without Evolution phase can also outperform Treefix, which reflects the effectiveness of the design for EnvAgent and RefAgent. (ii) EvoAgent brings benefits to the final performance with an up to around 22\% relative performance in FER. It also shows around 5\% increase in S-C and C-C. As these cases under EvoAgent are typically more difficult to execute than average, it could be hard to bring an extremely outstanding improvement.

\greyboxb{Summary of RQ3:~}{
We find that (i)~\name~has a diverse action space for various actions in different stages. (ii) Dynamic pruning improves the efficiency while increasing or maintaining the coverage performance in around 98.5\% cases. (iii) Prefix Evolution has positive effects on the results of RefAgent, with a relative improvement of up to 22\% in FER.
}

\section{Discussion}
\subsection{Case Study 1: Action Space}
\label{subsec:case1}
Here we study how~\name~executes one partial code snippet (\texttt{StackOverflow\_485}) with various actions. The following is the code that needs to be executed, and we omit some unnecessary parts about image processing: 

\setminted{
    breaklines,
    fontsize=\footnotesize,
    frame=single,
    tabsize=2,
    bgcolor=white
}
\begin{minted}{python}
# Importing Image class from PIL module
from PIL import Image

# Opens a image in RGB mode
im = Image.open(r"C:\Users\System-Pc\Desktop\ybear.jpg")

# [Omitted Image Processing Steps]
\end{minted}

\noindent To execute this code, there are two challenges: 
(i) There should be one valid image file for code to open and process, and (ii) the file system may not be compatible with the provided one (i.e., \texttt{r"C:\textbackslash Users\textbackslash ..."}). 
~\name~solves them in two different phases: 

(i) In the environment initialization stage, EnvAgent successfully identifies that the provided path of the image is Windows-style, and it is not compatible with its execution environment (i.e., Linux), shown in the THOUGHT of EnvAgent (where we omit unimportant texts for brevity): 

\begin{minted}[escapeinside=||]{text}
|\textcolor{brown}{\textbf{THOUGHT}}|: 
The code.py... opens a Windows-style path,... but the path is invalid in this Linux environment and the file is incomplete for execution... I will run the code... to use a local provided image or generate a placeholder image if the file doesn't exist...
\end{minted}

\noindent From the thought of the agent, we can also find that it would like to create an image file if there is no locally provided one (i.e., ``I will run the code... or generate a placeholder image...''). To realize this action, it utilizes its tool use capability and creates a Python script to check if one image exists or generates a simple one with the \texttt{Image} library: 

\begin{minted}[escapeinside=||]{python}
|\textcolor{brown}{\textbf{ACTION}}|:
python - << 'PY'
...
# We'll execute the logic in code.py by importing it if possible; otherwise, emulate running it safely.
# Since code.py is text, we will instead create a simple image and run a safe subset here to ensure PIL works.
img_path = Path("test_input_image.jpg")
if not img_path.exists():
    # create a simple image to crop
    img = Image.new("RGB", (400, 400), color=(123, 200, 150))
    ...
    img.save(img_path)
\end{minted}

(ii) After we already have one image for the script to use, in the second phase (i.e., Iterative Refinement), RefAgent attempts to solve the Windows-like path issue by mocking the used library \texttt{from PIL import Image}:

\begin{minted}{python}
...
# Define an Image "module" with an open function that returns MockImage
def mock_open(path):
    return MockImage(path)
# Create a module-like object for PIL and PIL.Image
pil_module = types.SimpleNamespace(
    Image=types.SimpleNamespace(open=mock_open))
# Ensure imports work: "from PIL import Image" will bind Image to pil_module.Image
sys.modules['PIL'] = pil_module
# Also provide 'PIL.Image' as a module entry
sys.modules['PIL.Image'] = pil_module.Image
\end{minted}

\noindent From the prefix~\name~appends, we can see that it uses \texttt{sys.modules} to mock the \texttt{Image} object used in the partial code; the method \texttt{Image.open} is also replaced by a self-defined function \texttt{mock\_open}. From this case, we can find that~\name~is able to create diverse contexts via the large action space powered by LLM agents, and consequently enable the execution of complicated partial code and achieve a full coverage (i.e., 1.0). In contrast, Treefix covers none of the lines of code in this case (i.e., 0.0).

\subsection{Case Study 2: Synthesizing Generator to Evolve Prefixes}
\label{subsec:case2}
We study~\name's actions for \texttt{\path{Open-Source_black_142}} as another example. In this case, the baseline Treefix achieves a 0.5 S-C, while~\name~increases the same metric from 0.69 (achieved in phase 2) to 0.88 through prefix evolution. The partial code is shown in the following code block:
\setminted{
    breaklines,
    fontsize=\footnotesize,
    frame=single,
    tabsize=2,
    bgcolor=white
}
\begin{minted}[escapeinside=||]{python}
has_value = leaf.type in BRACKETS or bool(leaf.value.strip())
if not has_value:
    exit()
if token.COLON == leaf.type and self.is_class_paren_empty:
    del self.leaves[-2:]
if self.leaves and not preformatted:
    leaf.prefix += whitespace(
        leaf, complex_subscript=self.is_complex_subscript(leaf)
    )
if self.inside_brackets or not preformatted or track_bracket:
    self.bracket_tracker.mark(leaf)
    if self.mode.magic_trailing_comma:
        if self.has_magic_trailing_comma(leaf):
            self.magic_trailing_comma = leaf
    elif self.has_magic_trailing_comma(leaf, ensure_removable=True):
        self.remove_trailing_comma()
\end{minted}

\noindent We can see that this code includes a lot of branches (e.g., \texttt{if} statements) and many of them are nested. In addition, it involves complex objects and attributes that influence each other (e.g., the object \texttt{leaf} and one attribute \texttt{\path{magic_trailing_comma}}). 
Therefore, it is difficult to achieve a high coverage on this code. 

In~\name, during the phase of iterative refinement, RefAgent successfully creates an executable prefix that contains complex elements and definitions (here we omit some statements for brevity): 
\begin{minted}[escapeinside=||]{python}
|\textcolor{brown}{\textbf{SEED PREFIX}}|:
class MockSelf:
    def __init__(self):
        self.mode = types.SimpleNamespace(magic_trailing_comma=False)
        self.inside_brackets = False
self = MockSelf()
leaf = Leaf(token.COLON, "X")
\end{minted} 
Though it has identified several important conditions to go into some deep branches (e.g., the deletion of \texttt{colon}), it fails to reach the nested logic code (i.e., the branches guarded by ``\texttt{if\\\ self.mode.magic\_trailing\_comma}'' and ``\texttt{has\_value}''). 
In our prefix evolution, the generator program iterates all the potential values of each variable and attribute (e.g., the parameter \texttt{True} or \texttt{False} in \\\texttt{self.mode = types.SimpleNamespace(magic\_trailing\_comma=)}), and successfully resolves this issue: 

\begin{minted}[escapeinside=||]{python}
|\textcolor{brown}{\textbf{EVOLVED PREFIX}}|:
class MockSelf:
    def __init__(self):
    self.mode = types.SimpleNamespace(magic_trailing_comma=True)
    self.inside_brackets = True
self = MockSelf()
leaf = Leaf(':', '')
\end{minted}

\subsection{Threats to Validity}
\phead{External Validity.} 
We conduct experiments with code sourced from real GitHub repositories and Stack Overflow questions in Python. Although they are adopted by previous works~\cite{souza2023lexecutor, souza2025treefix}, the choice of datasets and programming languages could still affect the results of our evaluation. 
We use the model GPT-5 mini/nano as the backbone model of~\name~for its agentic ability, compared with the GPT-4o series used by Treefix. To mitigate this issue, we reproduce the experiments of Treefix with the same base model and carefully ensure the fairness of the comparison by using the same experimental settings.
Mini-SWE-agent~\cite{mini_swe_agent} is selected as the agent scaffold of~\name. While it uses a general framework design (i.e., a minimal, ReAct-like agent), using other agent scaffolds may bring different conclusions.

\phead{Internal Validity.} 
One potential threat to internal validity is the design of prompts. We follow the prompt design protocol of mini-SWE-agent~\cite{mini_swe_agent} and adjust prompts based on preliminary experiments to improve the performance. To mitigate this issue, we open-source the full prompts we use in the experiments. 
In addition, LLM randomness may affect the evaluation results. Therefore, we disclose the source code with the specific API version to mitigate this issue. 
The environment provided for agents may also affect the internal validity. To lessen this, we use Dockerized~\cite{docker_website} containers to ensure a controlled execution environment. 

\phead{Construct Validity.}
One potential threat to construct validity is the choice of evaluation metrics. 
We use code coverage as the main metric to assess the quality of partial code execution. However, a higher coverage does not necessarily mean that this execution is better aligned with the real usage or intention. 
To mitigate this issue, we keep the evaluation metrics consistent with closely related works~\cite{xue2024selfpico,souza2025treefix,souza2023lexecutor}. As we use standard metrics used in prior studies, we believe the threat is minimal.

\vspace{-4pt}
\section{Conclusion and Future Work} 
In this paper, we study the problem of partial code execution, which aims to automatically execute incomplete code snippets. We first identify key limitations of previous works in their restricted action space, limited feedback mechanisms, and rigid optimization strategies. Then, we propose \name, a novel multi-agent framework that enables effective and efficient partial code execution. 
Extensive experiments on two popular datasets demonstrate that \name~significantly outperforms the state-of-the-art method, Treefix, in terms of both execution coverage and efficiency. 

\phead{Future work.} We plan to extend this work from various aspects. Agents often terminate prematurely, leaving trajectories incomplete and degrading downstream performance; customizing the harness to detect and recover from such early exits is a promising remedy. 
Besides, the granularity and diversity of data are also promising directions for future work.

\section*{Data Availability Statement}

Our code and data are publicly available at GitHub and Zenodo~\cite{replication}. We also include the appendix in the replication package. 

\section*{Acknowledgments}

This research/project is supported by the National
Research Foundation, Singapore, under its Investigatorship Grant (NRF-NRFI08-2022-0002). Any opinions, findings, and
conclusions or recommendations expressed in this
material are those of the author(s) and do not reflect the views of National Research Foundation,
Singapore.

\bibliographystyle{ACM-Reference-Format}
\bibliography{software}

\end{document}